\documentclass[amscd,amssymb,onecolumn,showpacs,verbatim,11pt]{revtex4}
\def\dspace{\baselineskip=0.3 in}
\usepackage{graphicx}
\usepackage{dcolumn}
\usepackage{bm}

\input epsf
\begin{document}

\dspace

\title{\Large NON LINEAR EQUATION OF STATE AND  EFFECTIVE PHANTOM DIVIDE IN  BRANE MODELS}
\smallskip

\author{\bf~Jibitesh~Dutta $^{1,3}$\footnote{jdutta29@gmail.com,~jibitesh@nehu.ac.in},
 Subenoy~Chakraborty $^2$\footnote{schakraborty@math.jdvu.ac.in}
and ~M.~Ansari $^3$\footnote{mansarinehu@gmail.com}}

\smallskip

\affiliation{$^{1}$Department of Basic Sciences and Social
Sciences,~ North Eastern Hill University,~NEHU Campus, Shillong -
793022 ( INDIA )}

\affiliation{$^2$Department of Mathematics,~Jadavpur
University,~Kolkata-32, ( INDIA )}

\affiliation{$^3$Department of Mathematics,~ North Eastern Hill
University,~NEHU Campus, Shillong - 793022 ( INDIA )}

\date{\today}

\begin{abstract}
Here, DGP model of brane-gravity is  analyzed and compared with
the  standard  general relativity and Randall-Sundrum cases using
non-linear equation of state. Phantom fluid is known to violate
the weak energy condition. In this paper, it is found that this
characteristic of phantom energy is affected drastically by the
negative brane-tension $\lambda$ of the RS-II model. It is found
that in DGP model strong energy condition(SEC) is always violated
and the universe accelerates only where as
in RS-II model  even SEC is not violated for $1 < \rho/\lambda < 2$ and the universe decelerates.\\\\\\\\\

\end{abstract}


\maketitle


\section{\normalsize\bf{Introduction}}

Astrophysical observations made during last few years have shown
conclusive evidence for cosmic acceleration at small
red-shift\cite{sp, sp03, ag}. It shows beginning of accelerate
expansion in the recent past. It is found that cosmic acceleration
is driven by some invisible fluid
 having its gravitational effect in the very late universe.
This unknown fluid has distinguishing feature of violating strong
energy condition(SEC) being called dark energy\cite{ejc}. Of late,
it was argued that observations are better fit for equation of
state parameter $\omega = p/\rho$ ($p$ being the pressure and
$\rho$ being the energy density) lying  in a narrow strip around
$\omega =-1$ and more towards  $\omega < -1$ leading to violation
of the weak energy condition (WEC) too\cite{rr, rr03}. This dark
energy fluid is popularly known as {\em phantom}. Using general
relativity(GR)
  based Friedmann equation, giving cosmic dynamics it is found that phantom dominated era
   of the universe accelerates but ends up in big-rip(BR) singularity after a finite time
   in future\cite{ejc}. Later on in \cite{sksjd} using brane-gravity(BG), singularity free phantom
   driven cosmic acceleration was obtained. Also, observations support homogeneous and flat model of the
universe \cite{ad}. The phantom model explains the present and
future acceleration of the universe, but it is
 plagued with the problem of big-rip singularity ( singularity in finite
 future time when energy density, pressure and the scale factor
 diverge). Thus, phantom  was another exotic matter suggested by Caldwell.
Different sources of exotic matter violating SEC \cite{jm, ca,
ttm, as,as1, mrg, psm, sk04, ms, gcs} and WEC were proposed in the
recent past \cite{rr, rr03, sk,  rj,ob,mc, sc, sc03, sm,add, mrw,
sno, sno03, abd,  om,ka, mb}. A comprehensive review of these
contributions is available in \cite{ejc}.

In the race to investigate a viable cosmological model, satisfying
observational constraints and explaining present cosmic
acceleration, brane-gravity was also drawn into service and
brane-cosmology was developed. A review on brane-gravity and its
various applications with special attention to cosmology is
available in\cite{var, rm,   pbx, cs}. In brane world scenario,
our four dimensional universe (a brane ) is a hypersurface
embedded in higher dimensional bulk spacetime. In this brane-bulk
scenario, all matter and gauge interactions (described by open
strings) are localised on a brane while gravity (described by
closed strings) may propagate into whole space time. This means
that gravity is fundamentally a higher dimensional interaction and
we only see the effective 4D theory on brane.

 Among the different brane proposals for brane models, the two
 prominent are RS (proposed by Randall and Sundrum) and DGP(proposed by
 Dvali, Gabadaze and Porrati) models. In RS models, the hierarchy
 problem could be solved by a warped or curved extra dimension
 showing that fundamental scale could be brought down from the
 Planck scale to 100 GeV. Thus, Randall-Sundrum approach brought the theory
 to scales below 100 GeV being the electroweak scale( so far results could be verified experimentally
upto this scale only). In this model, extra-dimension is large
having $(1 + 3)$-branes at its ends. These branes are
$Z_2$-symmetric (have mirror symmetry) and have tension to counter
the negative cosmological constant in the ``bulk'', which is ${\rm
AdS}_5$. The model, having {\em two} $(1 + 3)$-branes at the ends
of the orbifold $S^1/Z_2$, is known as RS-I model \cite{rs1}
whereas the model having  only {\em one} $(1 + 3)$-brane  is known
as RS-II model \cite{rs2}. In particular, RS-II model got much
attention due to its simple and rich conceptual base \cite{pb00,
pbd00, cg, cgs, mwb, cll, cal, apt,  sksgrg}. In case, the
extra-dimension is time-dependent, brane-gravity induced Friedmann
equation (giving dynamics of the universe) contains a correction
term $- {4\pi G
  \rho^2}/{3\lambda}$  with $\lambda$ being the
brane-tension \cite{var, rm,   pbx, cs} . In RS-I model, $\lambda$
is positive, whereas  $\lambda$ is negative in RS-II model.

 The main idea of the DGP model is the inclusion of a
 fourdimensional Ricci- Scalar into the action. On the
4-dimensional brane the action of gravity is proportional to
$M_{P}^{2}$ whereas in the bulk it is proportional to the
corresponding quantity in 5-dimensions. The model is then
characterized by  a cross over length scale
     $$ r_{c}=\frac{M_{P}^{2}}{2M_{5}^{2}}$$
such that gravity is 4-dimensional theory at scales $ a\ll r_{c}$
where matter behaves as pressure less dust but gravity
\textit{leaks out} into the bulk at scales $ a\gg r_{c} $ and
matter approaches the behaviour of a cosmological constant
\cite{dgp1,dgp2,dgp3}.

 In the standard cosmology RS gravity modifies the early universe
 where as DGP gravity modifies the late universe. So in the present
 universe brane corrections are not effective in RS-II model. But
 as phantom energy increases with the expansion, it brings drastic
 changes in the RS-II model with the equation of state (EOS) given by $p = -
\rho - f(\rho)$ with $f(\rho)$  being non-linear functional of
$\rho$.

 In our recent paper taking the above EOS we have analysed in
RS-II model and effective phantom divide is obtained in three
cases. So it is natural to study the effect of this EOS in DGP
model. The  aim of present paper is to extend the work of
\cite{sksjd} in DGP model and compare with corresponding solution
in RS-II model and GR cases. Further this EOS has been extensively
analysed by various authors in GR framework \cite{not, st, no}. So
it is imperative  to study  this EOS in Brane world scenario.
 This paper is organised as follows. In sec.2 we give the basic
 equations of GR, RS-II model and DGP model while in section 3, effective EOS is obtained
with brane-gravity corrections and contains discussion on
acceleration and deceleration of the models. Section 4 summarizes
the work.

 \section{\normalsize\bf{Basic Equations:}}
 Observations support homogeneous and isotropic model of the late universe,
given by the line-element \cite{ad}
 $$ ds^2 = dt^2-a^{2}(t)[dx^2 +dy^2 + dz^2]   \eqno(2.1)$$
where a(t) is the scale factor.

In this space-time the standard Friedmann equation in GR is given
by
$$ H^2 = \frac{\kappa^{2}}{3}\rho    \eqno(2.2)$$
where  $ \kappa^{2}= M_P^{-2}=8\pi G$

On the brane in both RS-II and DGP model the conservation equation
holds:
$$ {\dot \rho} + 3H(\rho + p) = 0 .    \eqno(2.3)$$

\subsection{\normalsize\bf{Randall-Sundrum II Brane World}}

In braneworld scenario, the gravity model is modified. In the
space-time given by eq(2.1), RS-II model based  Friedmann equation
is obtained as \cite{rm,  sksgrg}
$$ H^2 =  \frac{\kappa^{2}}{3}\rho\Big
[1-\frac{\rho}{2\lambda}\Big]    \eqno(2.4)$$ where $\lambda$ is
the negative brane tension.
\subsection{\normalsize\bf{Dvali-Gabadaze-Porrati Brane World}}

Unlike the RS-II model, the infinite extra dimension is flat $(
\Lambda_{5}=0)$ in DGP model. Brane tension is assumed to be zero
or cancelled out with a brane cosmological constant. In the
space-time given by eq(2.1), the DGP Friedmann equation is given
by\cite{dgp1,dgp2}
$$ H^{2}=\Big(\sqrt{\frac{\kappa^{2}\rho}{3}+\frac{1}{4r_{c}^{2}}}+\epsilon
\frac{1}{2r_{c}}\Big)^{2}       \eqno(2.5)$$
 or equivalently
$$ H^{2}-\epsilon \frac{H}{r_{c}}=\frac{\kappa^{2}\rho}{3}  \eqno(2.6)$$
 where $ H=\frac{\dot a}{a}$ is the Hubble parameter,~$\rho$ is the  cosmic fluid energy density and
  $ r_{c}=\frac{M_{P}^{2}}{2M_{5}^{2}}$ is the
crossover scale which determines the transition from 4D to 5D
behavior and $\epsilon = \pm 1 $.

For $ \epsilon = 1 $, we have standard DGP(+) model which is self
accelerating model without any form of dark energy, and effective
$\omega$ is always non phantom. However for $ \epsilon = - 1 $, we
have DGP(-) model which does not self accelerate but requires dark
energy on the brane. It experiences 5D gravitational modifications
to its dynamics which effectively screen dark energy.

In this paper we take DGP(-) model. Thus Friedmann eqs.(2.5) and
(2.6) simplifies to
$$
H^{2}=\Big(\sqrt{\frac{\kappa^{2}\rho}{3}+\frac{1}{4r_{c}^{2}}}-
\frac{1}{2r_{c}}\Big)^{2}       \eqno(2.7)$$ and
$$ H^{2}+\frac{H}{r_{c}}=\frac{\kappa^{2}\rho}{3}  \eqno(2.8)$$
\section{\normalsize\bf{Effective Equation of state and Cosmic Expansion :}}
Here the nonlinear equation of state(EOS) for phantom fluid is
taken as
$$p = -\rho - f(\rho)  \eqno(3.1)$$ where $f(\rho)=A
\rho^{\alpha}$ i.e.,
$$p = -\rho - A \rho^{\alpha}  \eqno(3.2)$$

Connecting eqs.(2.3) and (3.2) we have
$$\rho= \rho_{0}\Big[1+3 \tilde{A}(1-\alpha) ln\frac{a}{a_{0}}\Big]^{\frac{1}{1- \alpha}}
\eqno(3.3)$$where $ \alpha \neq 1$ and $\tilde{A}=A
\rho_{0}^{\alpha -1}$. This EOS is suitable for cosmological data
and centered around cosmological constant EOS $(A=0)$. Parameters
$A $  and $\alpha$ measures deviation from cosmological constant
EOS. The sign of the parameter A determines whether DE is phantom
or quintessence regime.

\subsection{\normalsize\bf{Randall-Sundrum II Brane World}}
Recently \cite{sksjd} we have analysed the EOS given by eq.(3.2)
in RS-II model. A part of the following discussion is based on
\cite{sksjd}.

Connecting Friedmann eq(2.2) and conservation equation
(2.3)\cite{nopl, sksgrg}we get
$$ \dot H + H^{2}=\frac{\ddot a}{a}=-\frac{\kappa^{2}}{2}(\rho
  +p)\Big[1-\frac{\rho}{\lambda}\Big]+\frac{\kappa^{2}}{6}\rho\Big[1-\frac{\rho}{2\lambda}\Big]
    \eqno(3.4)$$

  In GR-based theory, Friedmann equation is obtained as
$$\dot H + H^{2}=\frac{\ddot a}{a} = - \frac{\kappa^{2}}{3} [\rho + 3 P] . \eqno(3.5)$$

Comparing (3.4) and (3.5), the {\em effective EOS} with brane
gravity corrections is obtained as
$$ P = - \rho - f(\rho) \Big[1 - \frac{\rho}{\lambda}\Big] +
\frac{\rho^2}{3\lambda} \eqno(3.6)$$ using eq.(3.1).

Equation (3.6) yields
$$ \rho + P = - f(\rho)\Big[1 - \frac{\rho}{\lambda}\Big] +
\frac{\rho^2}{3\lambda} \eqno(3.7)$$ and
$$ \rho + 3 P = - 2\rho - 3 f(\rho) \Big[1 - \frac{\rho}{\lambda}\Big] +
\frac{\rho^2}{\lambda}   \eqno(3.8)$$ Here $ f(\rho)=A
\rho^{\alpha}$, therefore (3.6) yields
$$ P = - \rho -  A \rho^{\alpha}\Big[1 - \frac{\rho}{\lambda}\Big] +
\frac{\rho^2}{3\lambda} \eqno(3.9)$$

Equation (3.9) yields effective pressure $P < 0$ for
$$ \rho < 3 \lambda \Big[ 1 + A {\rho}^{({\alpha} - 1)}\Big\{1 -
\frac{\rho}{\lambda}\Big\}\Big]. \eqno(3.10)$$

In particular, if we choose $\alpha = 2$, then from eq(3.10)
energy density must restricted as

$$ \frac{3\lambda A -1 -\sqrt{D_{\lambda}}}{6A} < \rho<\frac{3\lambda A -1
+\sqrt{D_{\lambda}}}{6A}$$

where $D_{\lambda}=9\lambda^{2}A^{2}-42 A \lambda +1$ and assuming
$D_{\lambda}>0$

Further, it is found that
$$\rho + P = - A {\rho}^{\alpha} \Big[1 - \frac{\rho}{\lambda}\Big] +
\frac{\rho^2}{3\lambda} < 0     \eqno(3.11)$$ till
$$ \rho_0 <\rho < 3\lambda A {\rho}^{({\alpha} - 1)} \Big[1 -
\frac{\rho}{\lambda}\Big] \eqno(3.12)$$ with $ \rho_0$ being the
present energy density.

This result shows that WEC will be violated till $\rho$ will
satisfy the inequality (3.12). It will not be violated when
$$ \rho > 3\lambda A {\rho}^{({\alpha} - 1)} \Big[1 -
\frac{\rho}{\lambda}\Big]. \eqno(3.13)$$

This means that phantom fluid will behave effectively as phantom
dark energy till $\rho$ will obey the inequality (3.12). It will
not behave effectively as phantom when $\rho$  increases more and
 obeys the inequality (3.13).

Moreover, (3.8) shows that SEC will be  violated till
$$  \rho < \lambda \Big[2 +  3 A \rho^{(\alpha - 1)}
  \Big\{1-\frac{\rho}{\lambda}\Big\}  \Big]  .  \eqno(3.14)$$

This shows that when $\rho$ will increase such that
$$ 3\lambda A {\rho}^{({\alpha} - 1)} \Big[1 -
\frac{\rho}{\lambda}\Big] < \rho <
 \lambda \Big[2 +  3 A \rho^{(\alpha - 1)}
\Big\{1-\frac{\rho}{\lambda}\Big\}  \Big],   \eqno(3.15)$$ only
SEC will be violated. It shows that when $\rho$ will satisfy the
inequality (3.15), phantom
  characteristic to violate WEC will be suppressed by brane-gravity effects for
  negative brane tension and phantom fluid will behave effectively as
  quintessence. These
  results yield {\em effective phantom divide} at
$$\rho = \rho_{\rm phd} = 3\lambda A {\rho}_{\rm phd}^{({\alpha} - 1)} \Big[1 -
\frac{\rho_{\rm phd}}{\lambda}\Big]. \eqno(3.16)$$

It is interesting to see that even SEC will not be violated when
$$ \lambda \Big[2 +  3 A \rho^{(\alpha - 1)} \Big\{1-\frac{\rho}{\lambda}\Big\}
 \Big] < \rho < 2 \lambda. \eqno(3.17)$$
implying that, during the range (3.17), dark energy characteristic
to violate
  SEC and WEC will be suppressed completely by brane-corrections in RS-II model.

Eqs.(2.3),(2.4) and (3.2) yield
$$ \dot \rho - 3A\sqrt{\frac{\kappa^{2}}{3}}\rho^{\alpha
  +\frac{1}{2}}\sqrt{1-\frac{\rho}{2\lambda}}= 0 ,    \eqno(3.18) $$
where $\rho_0\leqslant\rho\leqslant 2\lambda .$

Exact solution of this equation is obtained as
 $$ t = \frac{1}{\sqrt{3}\kappa A}\Big[t_0 + 2{(2\lambda)}^{(1/2)-\alpha}
 \Big\{\sqrt{1-\frac{\rho_0}{2\lambda}} 2F_1\Big(\frac{1}{2},\frac{1}{2}+\alpha,
 \frac{3}{2},1-\frac{\rho_0}{2\lambda}\Big)$$
$$-\sqrt{1-\frac{\rho}{2\lambda}} 2F_1\Big(\frac{1}{2},\frac{1}{2}
+\alpha,\frac{3}{2},1-\frac{\rho}{2\lambda}\Big)\Big\}\Big] ,
\eqno(3.19) $$
  where $ 2F_1(a,b,c,x)$ is the hypergeometric function.Further, using (3.19) in (3.3),
  we get a relation between time $t$ and the scale factor $a(t)$.

As maximum value of $\rho$ is $2\lambda$, so phantom universe will
expand upto time $t_m$  given as
$$ t_m = \frac{1}{\sqrt{3}\kappa A}\Big[t_0 +
2{(2\lambda)}^{(1/2) -\alpha}\sqrt{1-\frac{\rho_0}{2\lambda}}
2F_1\Big(\frac{1}{2},\frac{1}{2}+\alpha,\frac{3}{2},1-\frac{\rho_0}{2\lambda}\Big)\Big]
\eqno(3.20)$$ with $t_0$ being the present time. Moreover, from
(3.3), it is obtained that
$$ 3(1-\alpha)Aln(a_m/a_0)={(2\lambda)}^{1-\alpha}-\rho_0^{1-\alpha},
\eqno(3.21)$$ where $ a_m = a(t_m).$ This equation shows that if
$\alpha \gtrless 1, 2\lambda >\rho_0$ as $a_m >a_0.$

From (3.16) and (3.19), we obtain {\em effective phantom divide}
at time
$$ t = t_{\rm phd} = \frac{1}{\sqrt{3}\kappa A}\Big[t_0 + 2{(2\lambda)}^{(1/2) -\alpha}
\Big\{\sqrt{1-\frac{\rho_0}{2\lambda}}
2F_1\Big(\frac{1}{2},\frac{1}{2}+\alpha,
\frac{3}{2},1-\frac{\rho_0}{2\lambda}\Big)$$
$$-\sqrt{1-\frac{\rho_{\rm phd}}{2\lambda}} 2F_1\Big(\frac{1}{2},\frac{1}{2}+\alpha,
\frac{3}{2},1-\frac{\rho_{\rm phd}}{2\lambda}\Big)\Big\}\Big] ,
\eqno(3.22) $$

Inequalities (3.15) and (3.17) show that for $\rho$ satisfying
$$  \rho <  \lambda \Big[2 +  3 A \rho^{(\alpha - 1)}
\Big\{1-\frac{\rho}{\lambda}\Big\}  \Big], \eqno(3.23)$$ SEC will
be violated. It means that the universe will accelerate till
$\rho$ will obey (3.23). But as $\rho$ will grow more and it will
satisfy
$$  \rho >  \lambda \Big[2 +  3 A \rho^{(\alpha - 1)}
 \Big\{1-\frac{\rho}{\lambda}\Big\}  \Big], \eqno(3.24)$$
SEC will not be violated. It means that the universe will
decelerate when
 $\rho$ will  obey the inequality (3.24). It shows a transition from
 acceleration to deceleration at $\rho  = \rho_{\rm tr}$ given by the equation
$$  \rho_{\rm tr} =  \lambda \Big[2 +  3 A \rho_{\rm tr}^{(\alpha - 1)}
 \Big\{1-\frac{\rho_{\rm tr}}{\lambda}\Big\}  \Big]. \eqno(3.25)$$

Connecting (3.19) and (3.25), we obtain that this transition will
take place at time
$$ t = t_{\rm tr} = \frac{1}{\sqrt{3}\kappa A}\Big[t_0 +
2{(2\lambda)}^{(1/2)
-\alpha}\Big\{\sqrt{1-\frac{\rho_0}{2\lambda}}
2F_1\Big(\frac{1}{2},\frac{1}{2}+\alpha,\frac{3}{2},1-\frac{\rho_0}{2\lambda}\Big)$$
$$-\sqrt{1-\frac{\rho_{\rm tr}}{2\lambda}}
  2F_1\Big(\frac{1}{2},\frac{1}{2}+\alpha,\frac{3}{2},1-\frac{\rho_{\rm
      tr}}{2\lambda}\Big)\Big\}\Big] . \eqno(3.26)$$

  \subsection{\normalsize\bf{Dvali-Gabadaze-Porrati Brane World}}

At high energy (early universe) $\frac{1}{r_{c}}$ is small and can
be neglected, therefore from eq(2.7), we obtain the standard GR
Friedmann equation as
$$ H^2 = \frac{\kappa^{2}}{3}\rho \eqno(3.27)   $$
In the late universe, the extra dimension effect cannot be
neglected and we shall use a approximation form of Friedmann
equation \cite{gum} as follows.

The DGP Friedmann equation (2.7) can be written as
$$ H^{2}=\frac{1}{4 r_{c}}\Big[\sqrt{1+\frac{4 \rho r_{c}^{2}}{3
M_P^2}} -1 \Big ]^2     \eqno(3.28)$$ Expanding in terms of $ \rho
r_{c}^{2}/M_P^2 \ll 1$, we get at the lowest order
$$H \approx\frac{\kappa^{2}r_{c}}{3} \rho   \eqno(3.29)$$

In this case, using (2.3) and (3.2), we have
$$ \dot H = \frac{\kappa^{4}r_{c}^{2} A}{3} \rho^{\alpha +1}  \eqno(3.30)$$
Consequently
$$ \dot H + H^{2}=\frac{\ddot
a}{a}=\frac{\kappa^{4}r_{c}^{2}}{3}\Big(A \rho^{\alpha
+1}+\frac{\rho^{2}}{3}\Big)    \eqno(3.31)$$ Comparing this
equation with GR based Friedmann equation (3.5), the effective EOS
in this case is obtained as
$$ P = - \rho -\frac{2}{3}\kappa^{2}r_{c}^{2}\rho^{2}\Big(\rho^{\alpha
-1}+\frac{1}{3}\Big)+\frac{2}{3}\rho    \eqno(3.32)$$

Equation (3.32) yields that
$$ \rho + P = \frac{2}{3}\rho -\frac{2}{3}\kappa^{2}r_{c}^{2}\Big(A \rho^{\alpha
+1} + \frac{\rho^2}{3}\Big)    \eqno(3.33) $$ and
$$ \rho + 3P = -2\kappa^{2}r_{c}^{2}\rho^2\Big(A \rho^{\alpha
-1} + \frac{1}{3}\Big)    \eqno(3.34)  $$

 From eq(3.34) we see that,unlike RS-II model , SEC is always violated in this
 case.Further from (3.33)
  $$\rho + P=\frac{2}{3}\rho -\frac{2}{3}\kappa^{2}r_{c}^{2}\Big(A \rho^{\alpha
+1} + \frac{\rho^2}{3}\Big) < 0 $$ till
$$ \rho_{0}< \rho < \kappa^{2}r_{c}^{2}\Big(A \rho^{\alpha
+1} + \frac{\rho^2}{3}\Big)    \eqno(3.35)$$ with $ \rho_{0}$
being the present energy density. In particular, if we choose
$\alpha = 2$, then for violation of WEC, the energy density must
be restricted as
$$ \rho>\frac{-1+3\sqrt{D_{r_{c}}}}{6A}$$

where  $D_{r_{c}}=\frac{1}{9}+\frac{4A}{\kappa^{2}r_{c}^{2}}$

The parameter $\tilde{A}$ in eq (3.3) determines the type of
behaviour of the DE density with the expansion of the universe.
The speed of change of the DE density is given by
$$\frac{d\rho}{da}=3\tilde{A}\rho_{0}\Big[1+3 \tilde{A}(1-\alpha) \ln\frac{a}{a_{0}}\Big]^{\frac{1}{1- \alpha}-1}
\eqno(3.36)$$ This shows that for $\tilde{A}>0$, DE density
grows,for $\tilde{A}= 0$,the dark energy is constant while for
$\tilde{A}<0$, it decreases with the expansion of the universe.
Thus for $\tilde{A}>0$, phantom energy increases with expansion
and there will be a time when $$\rho > \kappa^{2}r_{c}^{2}\Big(A
\rho^{\alpha +1} + \frac{\rho^2}{3}\Big)
  \eqno (3.37)$$
  For $\alpha = 2$, in this case energy density is restricted as
  $$ \rho<\frac{-1+3\sqrt{D_{r_{c}}}}{6A}$$

    This shows that WEC will be violated till $\rho$ satisfies the
  inequality (3.35). It will not be violated when $\rho$ satisfies
  (3.37). It means that when $\rho$ satisfies the equality (3.37)
  the phantom character  of violating WEC will be suppressed by
  brane-gravity effects and phantom fluid will behave effectively
  as quintessence. This yields the {\em effective phantom divide}
  at  $$\rho = \rho_{\rm phd} =\kappa^{2}r_{c}^{2}\Big(\tilde{A}\rho_{0}^{1-\alpha}
   \rho_{\rm  phd}^{\alpha
  +1}+\frac{\rho_{\rm phd}^{2}}{3}\Big)
 \eqno(3.38)$$
 Connecting MFE(modified Friedmann equation) (3.29) and conservation
 equation (2.3) we have
 $$ t = t_{0} - \frac{1}{\kappa^{2} \alpha \tilde{A}r_c \rho_{0}^{1-\alpha
}}\Big(\rho^{-\alpha}-\rho_{0}^{-\alpha}\Big)  \eqno(3.39) $$
Using eqs(3.38) and (3.39), we get the effective phantom divide at
time
$$ t_{\rm phd} = t_{0} - \frac{1}{\kappa^{2} \alpha \tilde{A}r_c \rho_{0}^{1-\alpha
}}\Big(\rho_{\rm phd}^{-\alpha}-\rho_{0}^{-\alpha}\Big)
\eqno(3.40)
$$ It is to be noted that in GR in order to have phantom crossing
EOS needs to have double value \cite{no}  i.e.,  $p = -\rho \pm
f(\rho)$

The eq. (3.39) can be written as

$$\rho= \rho_{0}\Big[1- \kappa^{2} \alpha \tilde{A}r_c \rho_{0}(t-
t_0)\Big]^{-\frac{1}{\alpha}}\eqno(3.41)
$$which can be rewritten as

$$ \rho =\frac{3 H_{0}^{2}\Omega_{0}}{\kappa^{2}} \Big[1-\frac{3\tilde{A}\alpha \Omega_{0}H_0}
{2c}(t-t_0)\Big]^{{-\frac{1}{\alpha}}} \eqno(3.42)$$ where
$$\Omega_{0}=\frac{ \kappa^{2} \rho_{0}}{3H_{0}^{2}}  \hspace{1.5cm} and
    \hspace{1.5cm} \Omega_{r_{c}}=\frac{1}{4r_{c}^{2}H_{0}^{2}}   \eqno(3.43)  $$
are the usual dimensionless density parameters.

In terms of these parameters Friedmann   eq (2.8) can be written
as
$$\Omega_{0}+2\sqrt{\Omega_{r_{c}}}=1   \eqno(3.44)$$

Solving (3.29) and (3.42) the scale factor in this case is
obtained as

$$a(t) = a_0 \exp \Big[\frac{1}{6 \tilde{A}(\alpha
-1)\sqrt{\Omega_{r_{c}}}}\Big(1-\frac{3\tilde{A}\alpha \Omega_{0}
H_0  (t-t_0)}{2
\sqrt{\Omega_{r_{c}}}}\Big)^{\frac{-1}{\alpha}}\Big(3\tilde{A}\alpha
\Omega_{0} H_0  (t-t_0)-2\sqrt{\Omega_{r_{c}}}\Big)
+2\sqrt{\Omega_{r_{c}}}\Big]  \eqno(3.45)$$

This is the expression of the scale factor at late stages of
evolution when $\rho$ is very small. Note that for $\alpha > 1$,
$a(t)$ grows exponentially with time while for $\alpha \leqslant
1$, $a(t)$ becomes constant asymptotically i.e., we shall have a
static model of the universe. Thus for $\alpha > 1$, the late
stage acceleration as demanded by the present day observation is
obtained while for $\alpha \leqslant 1$, the universe expands to a
(finite) maximum volume and then becomes static asymtotically.

\section{\normalsize\bf{Conclusions:}}
In this paper we analyze and compare behaviour of phantom fluid in
DGP model (normal branch) of brane gravity with RS-II model having
negative brane tension $lambda$. Brane corrections make drastic
changes to the behaviour of phantom fluid which is characterised
by violation of WEC and accelerating the universe  ending up in
big-rip singularity in most of the models.

In both DGP and RS-II model the phantom characteristic of
violating WEC is suppressed by brane gravity effects. Moreover
both models are free from big-rip problem. In RS-II model brane
tension plays a crucial role. When it is negative, it causes
drastic changes in the behaviour  of dark energy, but phantom
fluid has usual behaviour when brane tension is positive.

In DGP model SEC is always violated, consequently there is
acceleration only. Contrary to this  in RS-II model acceleration
is transient showing the non violation of even SEC. In RS-II model
the future universe begins at time $t_0$ , the present age of the
universe, and it stops expanding when energy density grows to $2
\lambda $ by time $t_m$. Contrary to this in DGP model there is no
maximum limit of expansion.\\\\
{\bf Acknowledgement:}\\
A part of the paper is done during a visit to IUCAA, Pune, India.
The first author is thankful to IUCAA for warm hospitality and
facility of doing
research works.\\\\

\end{document}